\documentclass[sigconf,authorversion]{acmart}

\usepackage{booktabs} 
\usepackage[export]{adjustbox} 
\usepackage{tikz}
\usetikzlibrary{positioning,calc,shadows,shadows.blur}
\usepackage{pgfplots}
\pgfplotsset{compat=1.6}
\usepackage{subcaption}
\usepackage{xcolor}
\usepackage{nicefrac}
\newcommand{\change}[1]{{#1}}

\usepackage[boxruled,lined]{algorithm2e} 
\SetCommentSty{emph}

\acmConference[I3D'22]{SIGGRAPH I3D}{May 2022}{} 
\acmYear{2022}
\copyrightyear{2022}

\acmDOI{10.1145/3522620}

\begin{document}
\title{Rendering Layered Materials with Diffuse Interfaces}

\author{Heloise de Dinechin}
\affiliation{%
  \institution{Unity Technologies, EPFL}
}
\orcid{0000-0002-6072-6167}  
\author{Laurent Belcour}
\affiliation{%
  \institution{Unity Technologies}
}
\orcid{0000-0002-1982-0717}

\begin{abstract}
In this work, we introduce a novel method to render, in real-time, Lambertian surfaces with a rough dieletric coating. We show that the appearance of such configurations is faithfully represented with two microfacet lobes accounting for direct and indirect interactions respectively. We numerically fit these lobes based on the first order directional statistics (energy, mean and variance) of light transport using 5D tables and narrow them down to 2D + 1D with analytical forms and dimension reduction. We demonstrate the quality of our method by efficiently rendering rough plastics and ceramics, closely matching ground truth. In addition, we improve a state-of-the-art layered material model to include Lambertian interfaces.
\end{abstract}

%
%


%
%


\begin{teaserfigure}
    \vspace{-10pt}
    {\begin{tikzpicture}[]
        \node[inner sep=0](img) {
            \adjincludegraphics[width=\linewidth,frame]{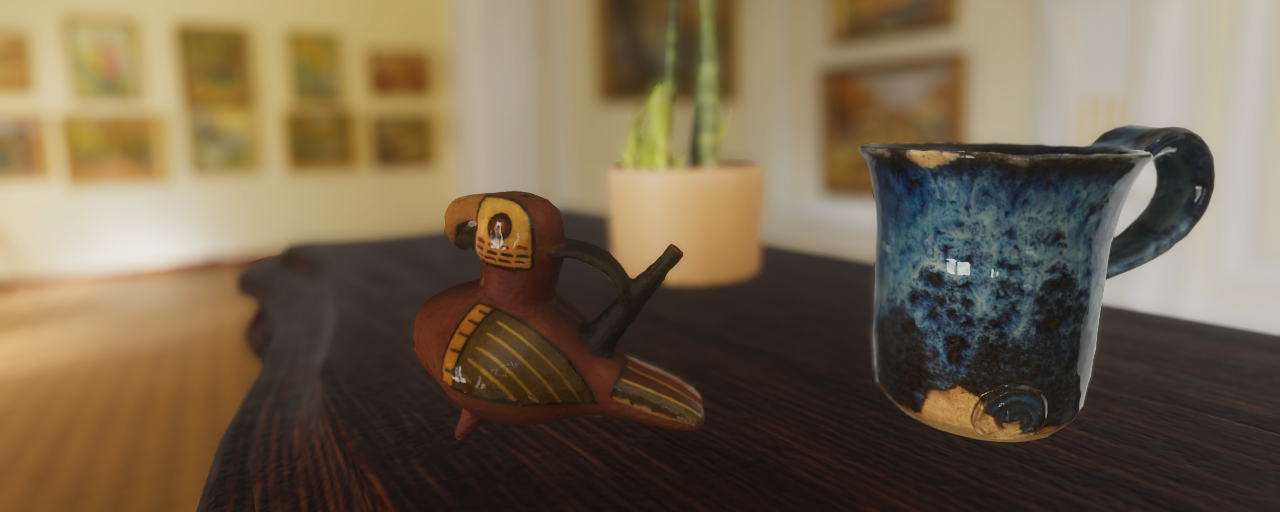}
        };
        \node[inner sep=0](box) {
            \adjincludegraphics[width=\linewidth,frame]{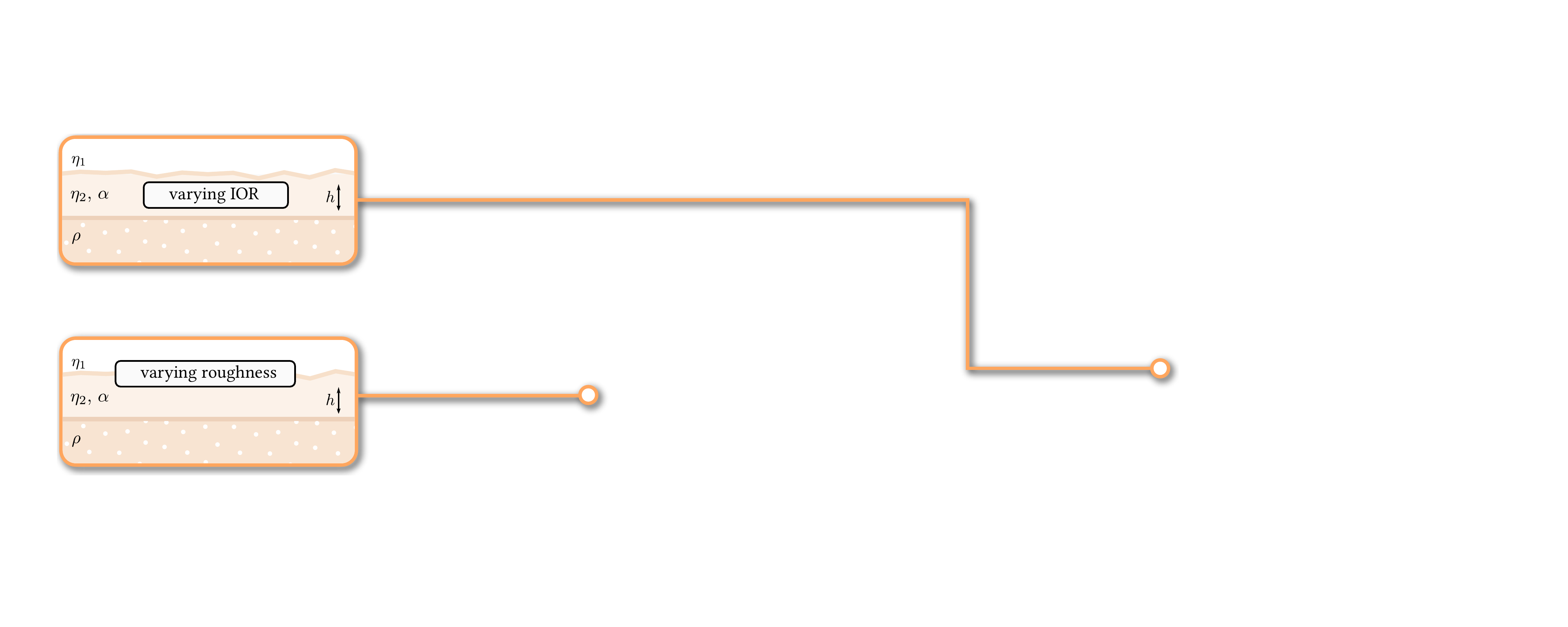}
        };
    \end{tikzpicture}
    \vspace{-15pt}}
    \caption{
    Real-time rendering of coated ceramic objects in the Unity game engine.
    \textmd{Our model enables efficient rendering of layered materials with
    a Lambertian base layer. Our method supports spatially varying parameters for albedo, coating roughness (here, for the parrot) index of refraction (here, for the mug), and Beer-Lambert attenuation.}
    \textmd{Models under CC-by scanned by Artur Wozniak and Jorgen Nelson}.
    \label{fig:teaser}
    }
\end{teaserfigure}

\maketitle

%
%
\section{Introduction}
\label{sec:introduction}

Material models that are both compatible with real-time and offline rendering engines are challenging to design. 
In many cases, real-time shading models crudely approximate offline rendering ones. An example is the case of coated ceramics or rough coated plastic, which are defined as a stack of a rough dieletric coating onto a Lambertian base. To approximate such materials, real-time models usually blend together a rough specular and a diffuse lobe using artist defined values, neglecting light transport. Offline models however can afford to evaluate the correct interaction between the coating and the base, achieving the correct saturation and brightness of the transmitted diffuse color, but at the cost of efficiency. \\

In this work, we provide a model to render such structures in real-time while being visually close to the ground truth. Our model builds on a simple idea. First, we tabulate the first order
directional statistics (energy, mean and variance) of both single and multiple scattering between the layers. Then, we use a sum of BRDF lobes which approximate those statistics. We use the fact that the Lambertian interface decorrelates the light transport integral to reduce those statistics to atomic precomputations combined by an analytical expression. Using BRDF lobes from a GGX microfacet model~\cite{walter2007} matching those statistics, we obtain an approximate model that closely resemble a stochastic reference. To better use the GPU's texture units, we compress the precomputed tables to 2D textures using an iterative dimensionality reduction. We further leverage this model to incorporate Lambertian interfaces into the statistical framework of Belcour~\shortcite{belcour2018}.

\begin{figure*}
    {\footnotesize
    \begin{tikzpicture}[]
        \node[inner sep=0](img) {
            \adjincludegraphics[width=\linewidth]{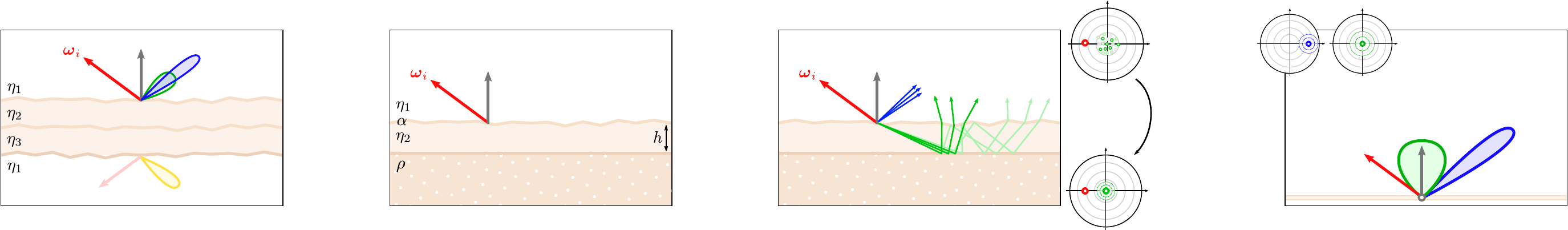}
        };
        \node at (-7.3, -1.3) {\textbf{(a) Layered Material}};
        \node at (-2.9, -1.3) {\textbf{(b) Rough Coated Lambertian}};
        \node at ( 1.5, -1.3) {\textbf{(c) Statistics Study}};
        \node at ( 7.2, -1.3) {\textbf{(d) Approximate Model}};
    \end{tikzpicture}
    \vspace{-20pt}
    }
    \caption{
        Rough coated Lambertian.
        \textmd{
            A layered material is defined as a stack of BSDF interfaces (a).
            We study a Lambertian coated by a rough dieletric microfacet interface (b). Such material is parameterized by its IOR $\eta = {\eta_1 \over \eta_2}$, roughness $\alpha$, albedo $\rho$, and unit-length absorption $\tau$.
            We study the statistics of light transport in such slab (c), and extract an approximate BRDF model (d) for direct and indirect lighting.
        }
        \label{fig:setup}
    }
\end{figure*}

\section{Previous Work}
Materials made of vertically stacked individual interfaces are called layered materials~\cite{weidlich2007} (Fig.~\ref{fig:setup}~(a)). The litterature can be split between methods that exhibit closed-forms models for a specific configuration, and frameworks made to combine an arbitrary number of layers.

\change{
\paragraph{Specific Models} provide closed-form expression for a specific number of layers.
For example, the Four-Flux Matrix method~\cite{simonot2016} specificaly models the transport between a dieletric coating and a Lambertian base. However, it does not provide a model compatible with prefiltered lights (such as HDRI or area-lights) that is mandatory for current real-time rendering usage.}

\paragraph{Offline Layered Materials Frameworks.}
The difficulty when designing generic frameworks resides in accounting for multiple scattering within the layered structure. Spectral representations~\cite{wjakob2014,zeltner2018} are efficient to evaluate such a model. There, the reflectance is decomposed in a Fourier basis and the adding-doubling algorithm~\cite{vandehulst80a} accounts for multiple scattering. \change{Unfortunately, this forbids the use of textures and requires a high evaluation overhead.}
Stochastic methods~\cite{guo2018,gamboa2020} use Monte Carlo method to evaluate the reflectance. They trade accuracy for noise and thus add variance to shading, thus restricting their use to offline scenarios.

\paragraph{Real-Time Layered Materials Frameworks.}
\change{Elek~\shortcite{Elek2010} adapted the model of Weidlich and Wilkie~\shortcite{weidlich2007} to real-time rendering. However, as with the original method, it lacks proper evaluation of multiple scattering and only support punctual light sources.}
Last, statistical models~\cite{guo2016,belcour2018} approximate the reflectance as a sum of microfacet lobes with directional albedo, incident direction and roughness reproducing the statistics of multiple scattering within the structure. \change{Thanks to this, it natively allows rendering extended light sources in real-time. Despite many extensions~\cite{yamaguchi2019,weier2020,randrianandrasana2021}, it does not handle Lambertian interfaces.}

\paragraph{Summary of Contribution} In this work, we use the statistical framework and extend it to handle Lambertian interfaces, making it possible to render a wider variety of configurations. To do so, we study the statistics of the stack of a rough dieletric coating on a Lambertian base (that we call a \textit{rough coated Lambertian}, Fig.~\ref{fig:setup}~(b)). From those statistics we both build a reflectance model for such material and improve the statistical layered framework to handle configuration with Lambertian bases (Fig.~\ref{fig:setup}~(c-d)).

%
%
\section{Statistics of a Coated Lambertian}
\label{sec:statistics}

\begin{figure*}[t]
    \pgfplotsset{
            compat=1.11,
            legend image code/.code={
                \draw[mark repeat=2,mark phase=2]
                plot coordinates {
                    (0cm,0cm)
                    (0.075cm,0cm)        
                    (0.15cm,0cm)         
                }   ;%
            }
        }
    \hspace{-10pt}
    \begin{subfigure}[b]{3.6cm}
        \begin{tikzpicture}[font=\footnotesize]
            \begin{axis}[
                grid=major,
                height = 4cm,
                xmin = 0.0, xmax = 1.0,
                ymin = 0.0, ymax = 1.0,
                xlabel = {elevation $\cos(\theta)$},
                legend style= {
                        at={(0.9, 0.65)},
                        nodes= {
                            scale=0.8,
                            transform shape
                        }
                    },
                ]
                \input{figures/tabulated/T01_a=000.txt}
                \addlegendentry{$\alpha = 0.00$}
                \input{figures/tabulated/T01_a=002.txt}
                \addlegendentry{$\alpha = 0.02$}
                \input{figures/tabulated/T01_a=006.txt}
                \addlegendentry{$\alpha = 0.06$}
                \input{figures/tabulated/T01_a=010.txt}
                \addlegendentry{$\alpha = 0.10$}
                \end{axis}
        \end{tikzpicture}
        \caption{$T_{01}$ for $\eta=1.5$}
    \end{subfigure}
    \begin{subfigure}[b]{4.9cm}
        \begin{tikzpicture}[font=\footnotesize]
            \begin{axis}[
                grid=major,
                name=energy,
                height = 4cm,
                xmin = 0.25, xmax = 4.0,
                ymin = 0.0, ymax = 1.0,
                xlabel = {IOR $\eta$},
                legend cell align=left,
                legend pos=outer north east,
                legend style= {
                        nodes= {
                            scale=0.8,
                            transform shape
                        }
                    },
                ]
                \addplot[black!0!cyan,line width=1.2pt] table {figures/tabulated/DT10_a=000.txt};
                \addlegendentry{$\bar{T}_{10}(\cdot, 0)$}
                \addplot[black!30!cyan,line width=1.2pt] table {figures/tabulated/DT10_a=050.txt};
                \addlegendentry{$\bar{T}_{10}(\cdot, {1\over 2})$}
                \addplot[black!50!cyan,line width=1.2pt] table {figures/tabulated/DT10_a=100.txt};
                \addlegendentry{$\bar{T}_{10}(\cdot, 1)$}

                \addplot[black!0!red,line width=1.2pt] table {figures/tabulated/DR10_a=000.txt};
                \addlegendentry{$\bar{R}_{10}(\cdot, 0)$}
                \addplot[black!30!red,line width=1.2pt] table {figures/tabulated/DR10_a=050.txt};
                \addlegendentry{$\bar{R}_{10}(\cdot, {1\over 2})$}
                \addplot[black!50!red,line width=1.2pt] table {figures/tabulated/DR10_a=100.txt};
                \addlegendentry{$\bar{R}_{10}(\cdot, 1)$}
            \end{axis}
        \end{tikzpicture}
        \caption{$\bar{T}_{10}$ and $\bar{R}_{10}$ for fixed $\alpha$}
    \end{subfigure}
    \begin{subfigure}[b]{5.3cm}
        \begin{tikzpicture}[font=\footnotesize]
            \begin{axis}[
                grid=major,
                name=energy,
                height = 4cm,
                xmin = 0.0, xmax = 1.0,
                ymin = 0.0, ymax = 1.0,
                xlabel = {elevation $\cos(\theta)$},
                legend cell align=left,
                legend pos=outer north east,
                legend style= {
                        nodes= {
                            scale=0.8,
                            transform shape
                        }
                    },
                ]
                \addplot[black!00!cyan,line width=1.2pt] table {figures/tabulated/EnergyEq_0000_0080.txt};
                \addlegendentry{$\rho_{2+}(\cdot, 0, 0.8)$}
                \addplot[black!30!cyan,line width=1.2pt] table {figures/tabulated/EnergyEq_0010_0080.txt};
                \addlegendentry{$\rho_{2+}(\cdot, 0.1, 0.8)$}
                \addplot[black!50!cyan,line width=1.2pt] table {figures/tabulated/EnergyEq_0050_0080.txt};
                \addlegendentry{$\rho_{2+}(\cdot, 0.5, 0.8)$}
                
                \addplot[black!00!red,line width=1.2pt] table {figures/tabulated/EnergyEq_0000_0250.txt};
                \addlegendentry{$\rho_{2+}(\cdot, 0, 2.5)$}
                \addplot[black!30!red,line width=1.2pt] table {figures/tabulated/EnergyEq_0010_0250.txt};
                \addlegendentry{$\rho_{2+}(\cdot, 0.1, 2.5)$}
                \addplot[black!50!red,line width=1.2pt] table {figures/tabulated/EnergyEq_0050_0250.txt};
                \addlegendentry{$\rho_{2+}(\cdot, 0.5, 2.5)$}

            \end{axis}
        \end{tikzpicture}
        \caption{Directional Energy $\rho_2+(\cos(\theta), \eta, \alpha)$}
    \end{subfigure}
    \begin{subfigure}[b]{4cm}
        \begin{tikzpicture}[font=\footnotesize]
            \begin{axis}[
                grid=major,
                name=energy,
                height = 4cm,
                xmin = 0.25, xmax = 4.0,
                ymin = 0.0, ymax = 0.3,
                xlabel = {IOR $\eta$},
                legend pos=south east,
                legend style= {
                        nodes= {
                            scale=0.8,
                            transform shape
                        }
                    },
                ]
                \addplot[black!0!red,line width=1.2pt] table {figures/tabulated/Var_a=000.txt};
                \addlegendentry{$\alpha = 0.0$}
                \addplot[black!20!red,line width=1.2pt] table {figures/tabulated/Var_a=010.txt};
                \addlegendentry{$\alpha = 0.1$}
                \addplot[black!40!red,line width=1.2pt] table {figures/tabulated/Var_a=030.txt};
                \addlegendentry{$\alpha = 0.3$}
                \addplot[black!60!red,line width=1.2pt] table {figures/tabulated/Var_a=050.txt};
                \addlegendentry{$\alpha = 0.5$}
            \end{axis}
        \end{tikzpicture}
        \caption{Variance $\sigma_{2+}\left(\eta, \alpha\right)$}
    \end{subfigure}
    \caption{Statistics of a Coated Lambertian. \textmd{From the precomputed rough transmission $T_{01}$ (a), precomputed diffuse reflection $\bar{R}_{10}$ and precomputed transmission $\bar{T}_{10}$ (b), we can compute the the energy of multiple scattering using Equation~\ref{eqn:directional_energy} and obtain the direction albedo (c). Similarly, we precompute its variance in the projected disc of outgoing directions (d).}
    \label{fig:tabulated}
    }
\end{figure*}

The \textit{Rough Coated Lambertian} reflectance is the sum of light directly reflected by the coating and light indirectly reflected by the Lambertian base. We opted to decompose the resulting BRDF model in two terms: a direct term $\rho_1$ and an indirect term $\rho_{2+}$. The direct term is already described in previous work~\cite{belcour2018}. For the indirect term, we study its energy, mean and variance.

\subsection{Energy}
The energy for the multiple scattering component $\rho_{2+}$ is the integral of paths that interact once with the Lambertian base. They undergo the rough refraction $T_{01}$ from $\eta_1$ to $\eta_2$, the diffuse interaction $\rho(x)$, and the rough reflection and refraction from $\eta_2$ to $\eta_1$, $\bar{R}_{10}$ and $\bar{T}_{10}$ :
\begin{align}
    \rho_{2+} &= \sum_{\Omega_n} \left( \int_{\mathbf{x} \in \Omega_n} T_{01}(x_0) \rho(x_1) \left[ \prod_{k=2}^{n-2} \bar{R}_{10}(x_k) \rho(x_{k+1}) \right] \bar{T}_{10}(x_{n}) \right) \nonumber
\end{align}
where $\mathbf{x} = {x_i}_{[0,n]} \in \Omega_n$ is all paths of length $n$.
Because of the Lambertian interaction, the integral at each vertex are uncorrelated:
\begin{align}
    \rho_{2+} &= \sum_{n} T_{01} \, \rho \, \left[ \prod_{k=2}^{n-2} \bar{R}_{10} \, \rho \right] \bar{T}_{10}\\
    &= \sum_{k=0}^\infty T_{01} \, \rho^{k+1} \, \bar{R}_{10}^k \, \bar{T}_{10}
\end{align}
where $T_{01} = \int T_{01}(\mathbf{x}) \mbox{d}\mathbf{x}$. This form is a convergent series:
\begin{align}
    \rho_{2+}  &= T_{01} \, \dfrac{\rho}{1 - \rho \, \bar{R}_{10}} \, \bar{T}_{10}. \label{eqn:directional_energy}
\end{align}
Where $\bar{T}_{01}$ depends on $\eta, \alpha$ and the incident angle, $\bar{R}_{10}$ and $\bar{T}_{10}$ depends on $\eta$ and $\alpha$ (see Figure~\ref{fig:tabulated}~(a-c)). Therefore, this expresses the energy with 3D and 2D function instead of a 4D one.

\subsection{Mean}
The mean of the indirect lobe is always the shading normal. Indeed, the Lambertian interaction distributes energy symmmetrically in the hemisphere, around the shading normal. The rough transmission and rough reflection afterwards do not change such symmetry.

\subsection{Variance}
\label{sec:statistics_variance}
We used a virtual goniophotometer to record the directional variance in the projected tangent plane. We traced paths in the coated lambertian structure by initiating rays from the diffuse base as light distribution is not impacted by incident directions. Hence we obtained a 2D table $\sigma_{2+}(\eta, \alpha)$, that we display in Figure~\ref{fig:tabulated}~(d).

\subsection{An Approximate Model}
Using this data, an approximate rough coated Lambertian model can be instanciated as the sum of two BRDF lobes: the classic microfacet model $\mbox{BRDF}_1$ accounting only for the reflection by the GGX distribution of normals of roughness $\alpha$ and of IOR $\eta$, and the indirect term $\mbox{BRDF}_{2+}$:
\begin{align}
    \mbox{BRDF}\left( \omega_i, \omega_o \right) = \mbox{BRDF}_1\left( \omega_i, \omega_o \right) + \mbox{BRDF}_{2+}\left( \omega_i, \omega_o \right),
    \label{eqn:approximate_model}
\end{align}
with the second term as:
\begin{align}
    {BRDF}_{2+}\left( \omega_i, \omega_o \right) = \rho_{2+}  \; \dfrac{D(\mathbf{n}, \omega_o, \alpha_{2+}) \bar{G}(\mathbf{n}, \omega_o, \alpha_{2+})}{4 \cos(\theta_i) \cos(\theta_o)},
\end{align}
where $D(\cdot, \cdot)$ is the microfacet distribution, $\bar{G}(\cdot, \cdot)$ is a \textit{normalized} shadowing/masking term (since energy loss due to high roughnesses is accounted in $\rho_{2+}$), and $\alpha_{2+}$ is the equivalent roughness for the variance calculated in Section~\ref{sec:statistics_variance}. Notice how the normal distribution and the shadowing/masking term are evaluated with the shading normal as the incident direction. \\

\begin{figure}[b]
    { \footnotesize
    \begin{tikzpicture}[]
        \node[inner sep=0] (gl) {
            \adjincludegraphics[height=3cm, frame, clip, trim=164 0 640 0]{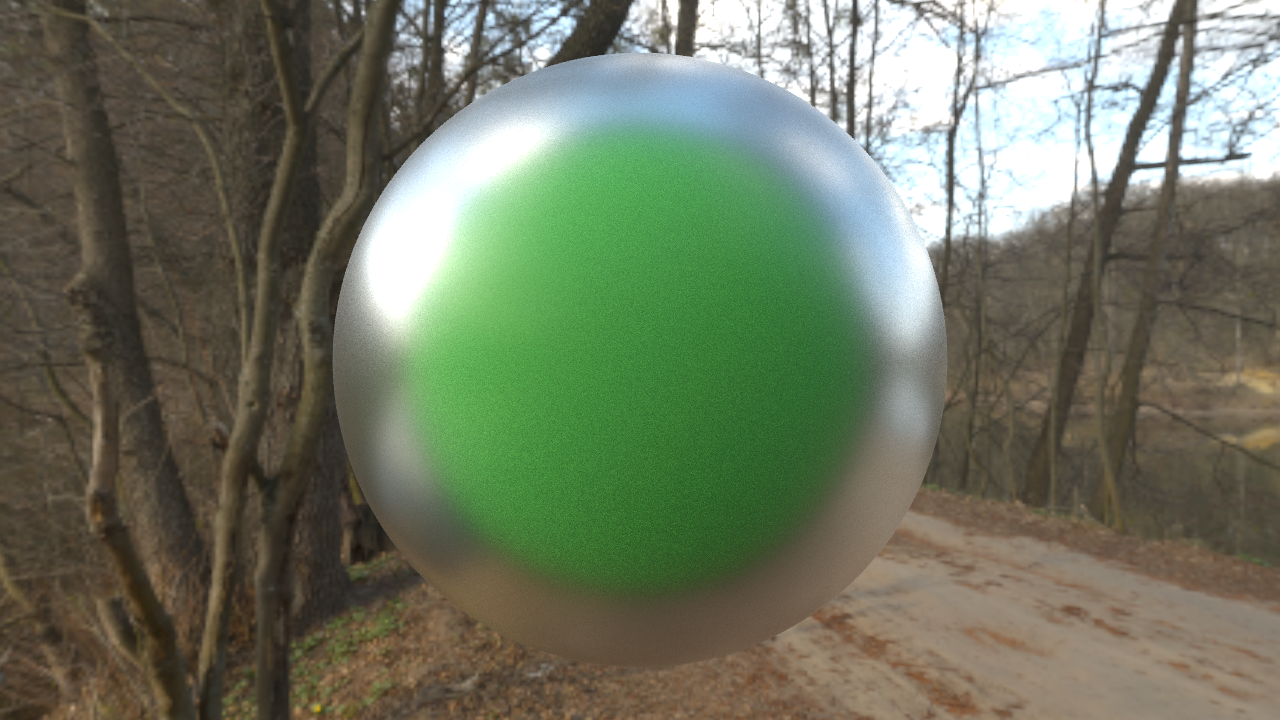}
        };
        \node[below=-1pt of gl, anchor=south] {\color{white} Ours (OpenGL) };
        \node[inner sep=0, right=-1pt of gl, anchor=west] (ref) {
            \adjincludegraphics[height=3cm, frame, clip, trim=640 0 164 0]{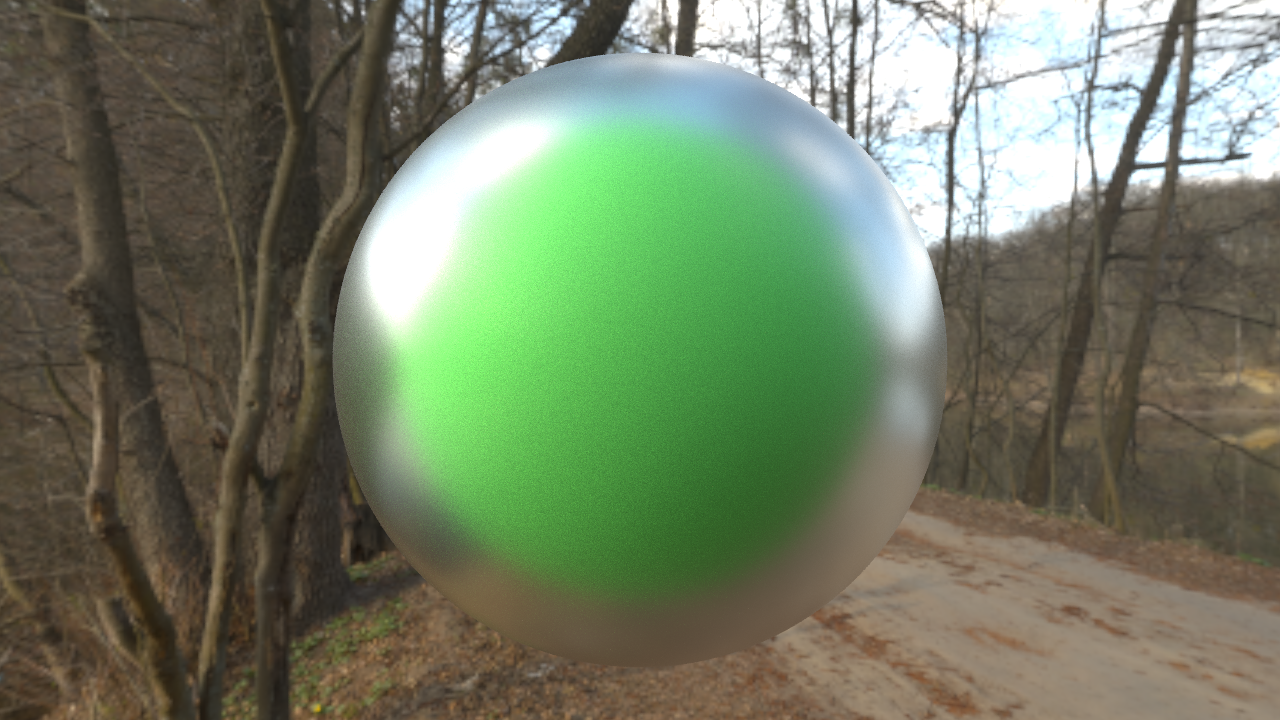}
        };
        \node[below=-2pt of ref, anchor=south] {\color{white} Reference };
        \node at (1.0, -1.8) { $\eta = 0.8, \, \alpha = 0.1, \, \rho =\left[ 0.2, 1.0, 0.2\right]$};
    \end{tikzpicture}
    \begin{tikzpicture}[]
        \node[inner sep=0] (gl) {
            \adjincludegraphics[height=3cm, frame, clip, trim=164 0 640 0]{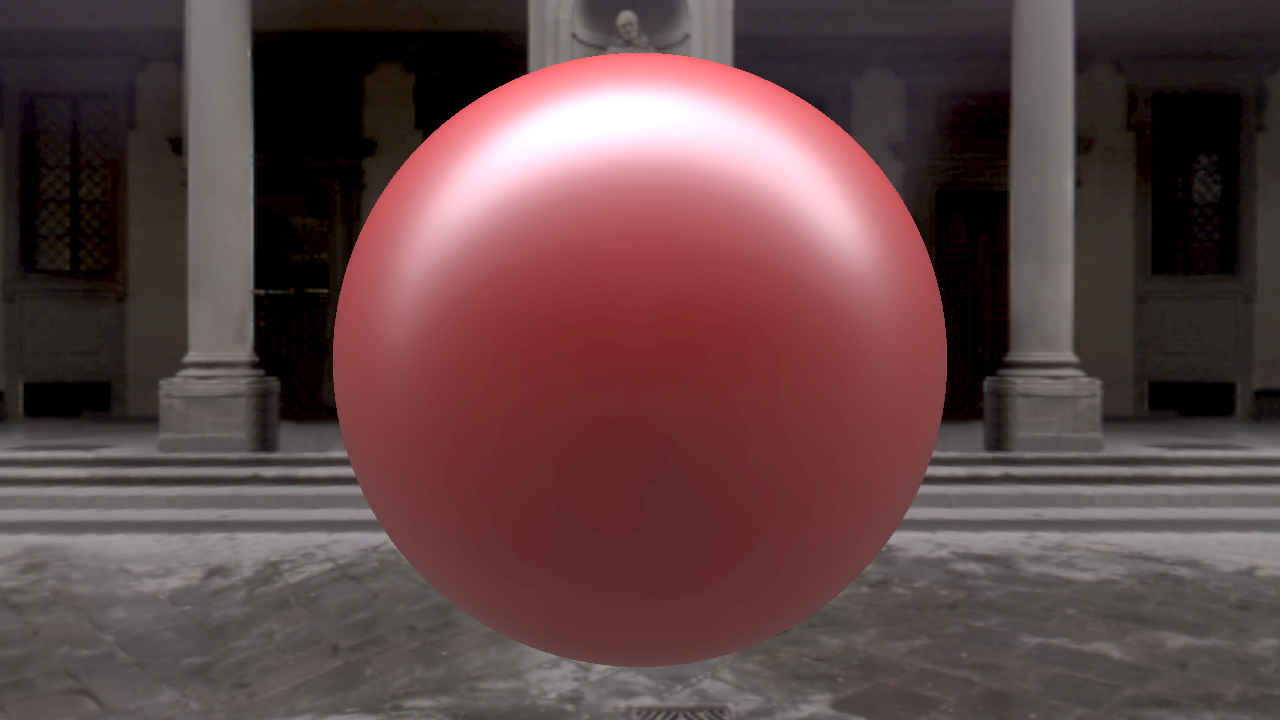}
        };
        \node[below=-1pt of gl, anchor=south] {\color{white} Ours (OpenGL) };
        \node[inner sep=0, right=-1pt of gl, anchor=west] (ref) {
            \adjincludegraphics[height=3cm, frame, clip, trim=640 0 164 0]{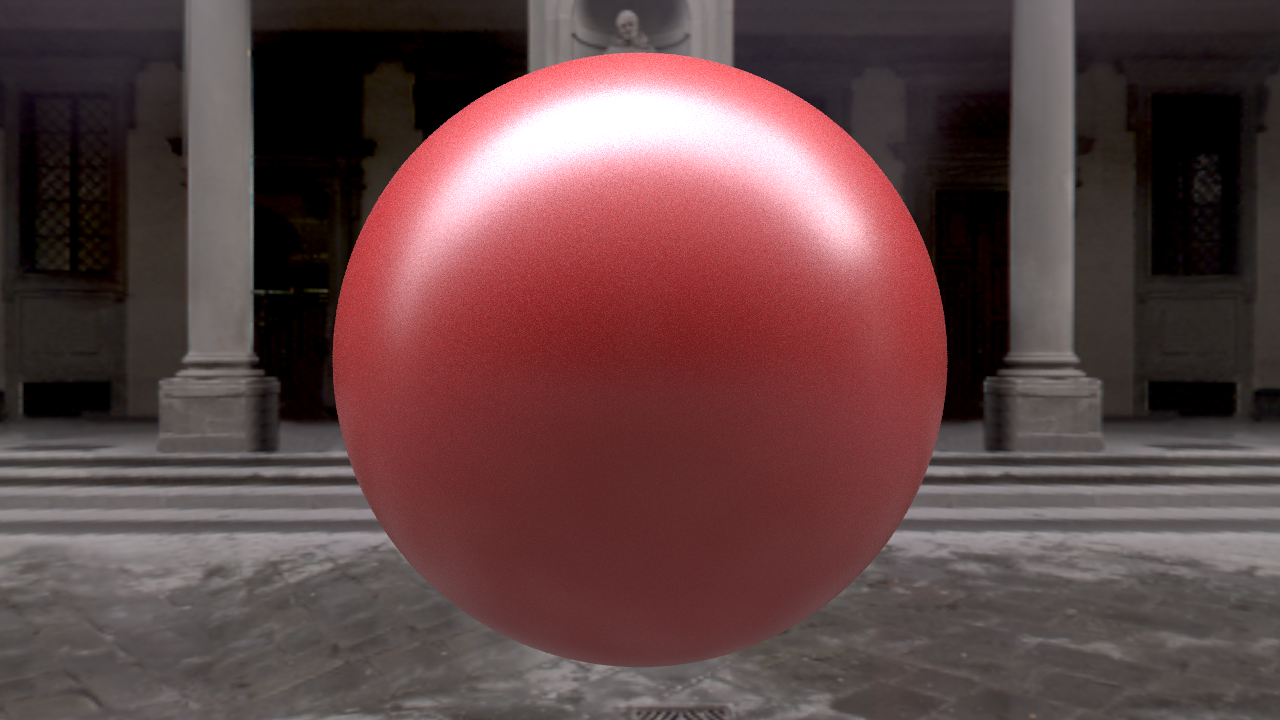}
        };
        \node[below=-2pt of ref, anchor=south] {\color{white} Reference };
        \node at (1.0, -1.8) { $\eta = 1.2, \, \alpha = 0.1, \, \rho =\left[ 1, 0.1, 0.1\right]$};
    \end{tikzpicture}
    }
    \vspace{-10pt}
    \caption{
        An approximate model for rough coated Lambertian.
        \textmd{
            We compare our Real-time approximation of Equation~\ref{eqn:approximate_model} to a stochastic reference in Mitsuba. We resemble the reference for various configuration either with $\eta < 1$ (left) or $\eta > 1$ (right) configurations. See our supplemental material for more results.
        }
        \label{fig:approximate_model}
    }
\end{figure}

We compare this real-time model to an offline reference in Figure~\ref{fig:approximate_model}. Our work matches difficult configurations such as $\eta < 1$, where total internal reflection generates moving discontinuities of varying smoothness depending on the roughness of the dielectric interface. Those configurations need to be accounted for if we want to add more layers on top or render underwater scenes.

%
%
\section{Reducing the Dimension of Precomputed Tables}
\begin{figure}[b]
    { \footnotesize
    \begin{tikzpicture}[font=\footnotesize]
        \begin{scope}
            \node[inner sep=0] (gl) {
                \adjincludegraphics[height=2.5cm, frame, clip, trim=140 0 640 0]{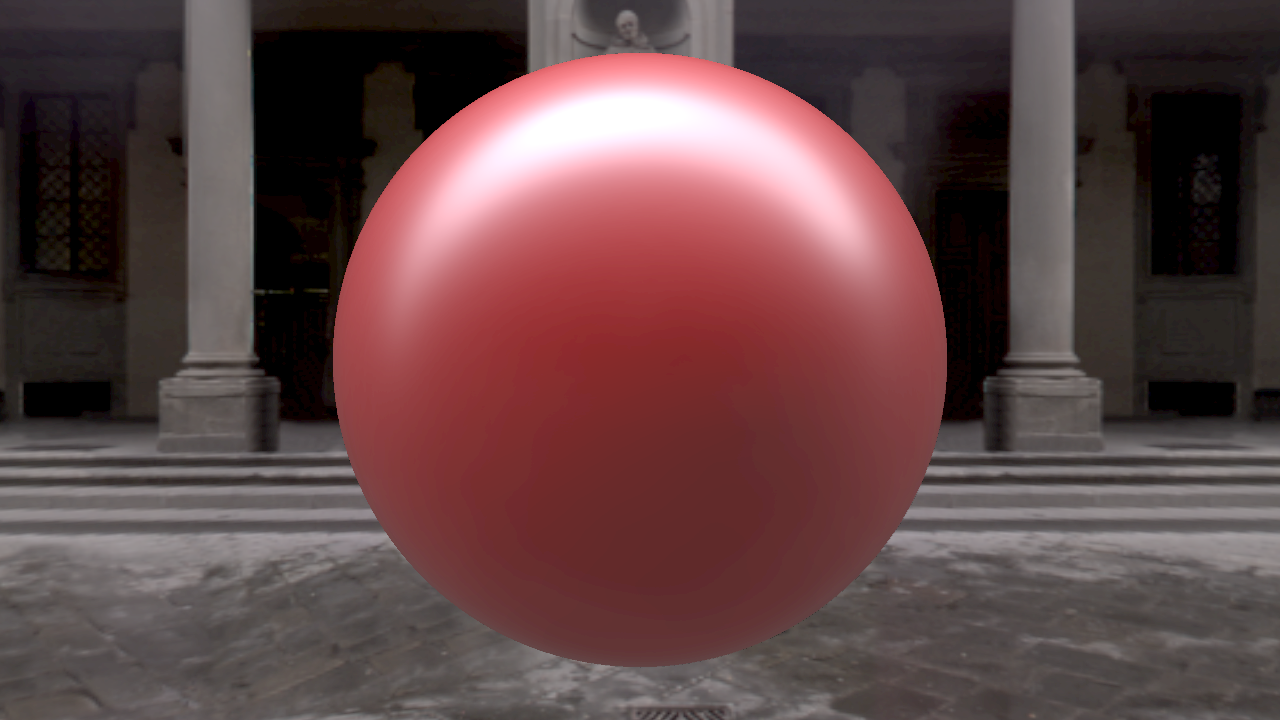}
            };
            \node[below=-2pt of gl, anchor=south] {\color{white} 4D };
            \node[inner sep=0, right=-1pt of gl, anchor=west] (ref) {
                \adjincludegraphics[height=2.5cm, frame, clip, trim=640 0 140 0]{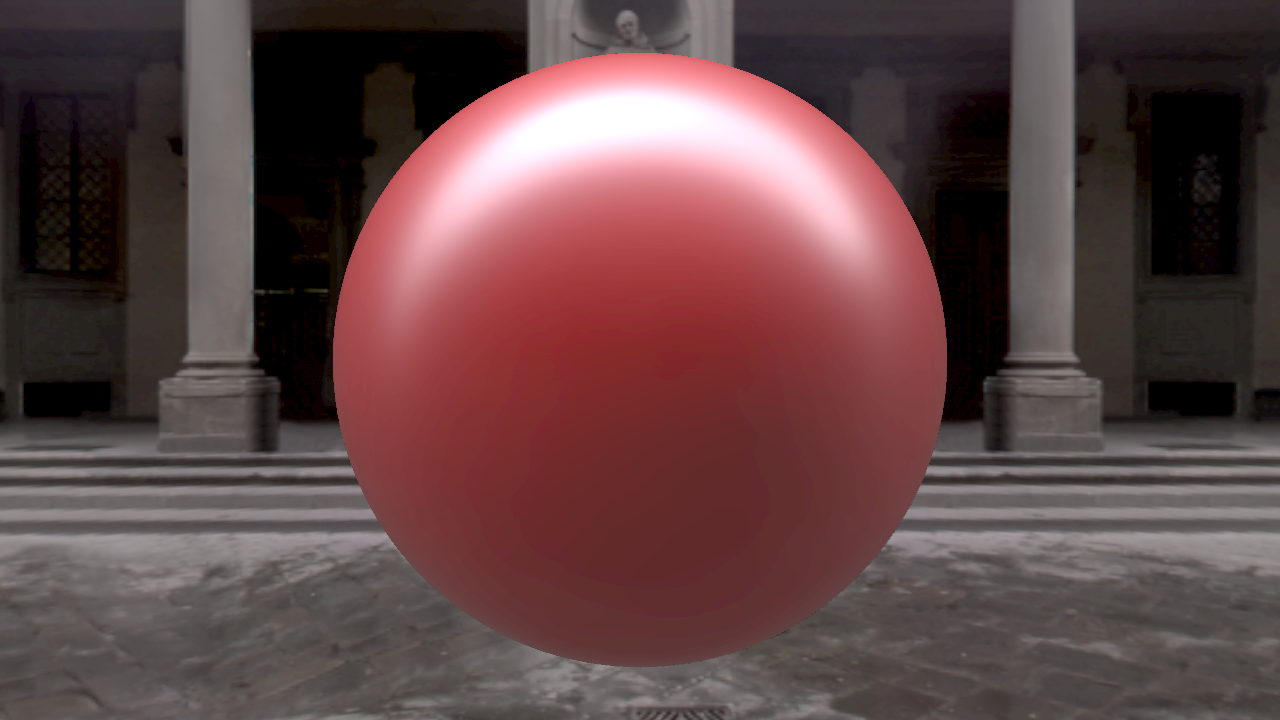}
            };
            \node[below=-2pt of ref, anchor=south] {\color{white} $2D+1D+1D$ };
            \node at (1.0, -1.8) { $\eta = 1.2, \, \alpha = 0.1, \, \rho =\left[ 1, 0.1, 0.1\right]$};
        \end{scope}

        \begin{scope}[shift={(3.6,-1.25)}]
            \begin{axis}[
                grid=major,
                name=energy,
                height = 4cm,
                xmin = 0.0, xmax = 1.0,
                ymin = 0.0, ymax = 1.0,
                xlabel = {$\cos\left( \theta \right)$},
                ylabel = {$T_{01}(\cdot, \eta=2, \alpha=0, \tau=1)$},
                legend pos=south east,
                legend style= {
                        nodes= {
                            scale=0.8,
                            transform shape
                        }
                    },
                ]
                \addplot[black!0!red,line width=1.2pt] table {figures/dimensions/plots/T01_200_0_100.txt};
                \addlegendentry{raw data}
                \addplot[dashed, black!20!red,line width=1.2pt] table {figures/dimensions/plots/T01_reconstructed_200_0_100.txt};
                \addlegendentry{compressed}
            \end{axis}
        \end{scope}
    \end{tikzpicture}
    }
    \vspace{-10pt}
    \caption{
        Data Compression.
        \textmd{
            We compress $T_{01}$ from $4D$ to $2D + 1D + 1D$ tables, $\bar{T}_{10}$ and $\bar{R}_{10}$ from 3D to $2D + 1D$  using 2 basis for $\tau$ and 4 basis for $\alpha$. This results in small rendering differences. 
        }
        \label{fig:dimensions}
    }
\end{figure}

Our model handles Beer-Lambert extinction by adding $\tau = \exp\left(- \sigma_a\right)$ as an additional dimension for $T_{01}$, $\bar{T}_{10}$ and $\bar{R}_{10}$. However, storing a dense 4D table for $T_{01}$ becomes prohibitive on GPU. We reduce the dimensionality of the data by applying a Principal Componnent Analysis (PCA) and decomposing it into a 1D basis table and a 3D coefficients one:
\begin{align}
    T_{01}\left( \cos(\theta_i), \alpha, \eta, \tau \right) \simeq \sum_{k=0}^{N} c_k\left(\cos(\theta_i), \alpha, \eta\right) b_k(\tau).
\end{align}
The coefficients $c_k\left(\cos(\theta_i), \alpha, \eta\right)$ can be further broken down into a similar decomposition:
\begin{align}
    c_k\left( \cos(\theta_i), \alpha, \eta \right) \simeq \sum_{j=0}^{M} c^\prime_j\left(\cos(\theta_i), \alpha\right) b^\prime_j(\eta),
\end{align}
which gives:
\begin{align}
    T_{01}\left( \cos(\theta_i), \alpha, \eta, \tau \right) \simeq \sum_{k=0}^{N} \sum_{j=0}^{M} c^\prime_j\left(\cos(\theta_i), \alpha\right) b^\prime_j(\eta) b_k(\tau).
\end{align}
Applying this strategy multiple times allows us to reduce the dimensionality of the tables. When the variations for the last dimensions are low frequency, the required number of basis components is small and can be stored in a single texture (using 4 basis for example). We show the result of this compression in Figure~\ref{fig:dimensions}.

%
%
\section{Statistical Layered Framework}
\begin{figure}[h!]
    { \footnotesize
    \begin{tikzpicture}[]
        \node[inner sep=0] (framework) {
            \adjincludegraphics[width=\linewidth]{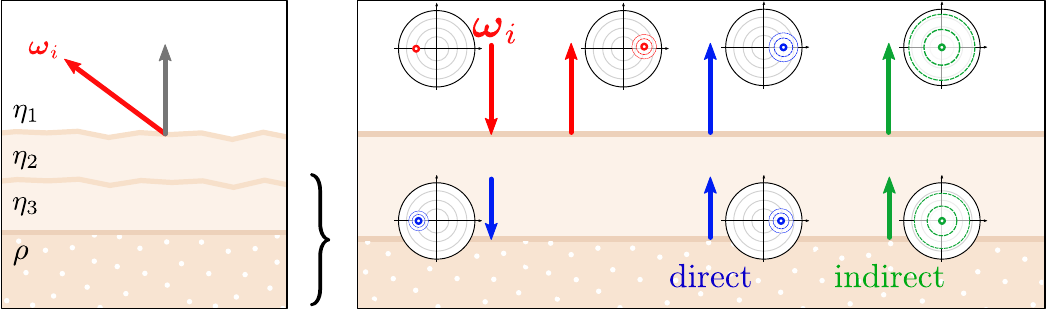}
        };
    \end{tikzpicture}
    }
    \vspace{-20pt}
    \caption{
        Updating the layered framework.
        \textmd{
            We add the rough coated Lambertian as a new interface in the statistical layered framework.
            We adapt the adding-doubling algorithm to track the statistics of a new lobe centered around the shading normal (green). 
        }
        \label{fig:update_layered_framework}
    }
    \vspace{-20pt}
\end{figure}

Thanks to the statistics gathered in Section~\ref{sec:statistics}, we add support for Lambertian interfaces in the statistical layered framework~\cite{belcour2018}. \change{For that, we group together the last dieletric and Lambertian interfaces to form a single layer interface. Contrary to other layers, this new one outputs two BRDF lobes. To propagate this additional lobe, we add another set of directional statistics to the adding-doubling algorithm (Figure~\ref{fig:update_layered_framework}) and update it (Algorithm~\ref{alg:new_adding_doubling}).}

\vspace{-2pt}

\change{
\paragraph{The adding-doubling algorithm} Inside the statistical framework, operators are used to describe the change of the directional moments (energy $e$, mean $\mu$, and variance $\sigma$) when light is either reflected or refracted by a layer. Those \textit{local} statistics are used to update \textit{global} statistics such as the transmission and reflection coefficients for light entering from above the surface: $t_{0i}$ and $r_{0i}$, and transmission and reflection coefficients for light entering from below the surface: $t_{i0}$ and $r_{i0}$. All those coefficients are computed during the adding-doubling algorithm and are combined to evaluate the different lobes approximating the material's reflectance.
}

\begin{algorithm}[t!]
    \caption{
        \textbf{Updated Adding-Doubling Algorithm.}
        Adding a Lambertian base to a dieletric layer stack requires a bottom-to-top evaluation loop to account for energy scale, multiple scattering and lobe scaling.
        We highlight the changes we made to the original algorithm in red.
        \label{alg:new_adding_doubling}
    }
    
    $r_{0i} = 0; r_{i0} = 0$\tcp*[r]{\color{green!50!black} Reflectances for $i$ layers}
    $t_{0i} = 1; \; t_{i0} = 1$\tcp*{\color{green!50!black} Transmittances for $i$ layers}
    $\sigma^{R}_{0i} = 0; \sigma^{R}_{i0} = 0$\tcp*[r]{\color{green!50!black} Variance of reflectances for $i$ layers}
    $\sigma^{T}_{0i} = 0; \sigma^{T}_{i0} = 0$\tcp*[r]{\color{green!50!black} Variance of transmittances for $i$ layers}
    $j_{i0} = 1$\tcp*{\color{green!50!black} Jacobian from i\textsuperscript{th} to top layer}
    
    \BlankLine
    // {\color{green!50!black} \emph{Top-to-Bottom loop to evaluate dieletric lobes} } \\
    \For{\color{red!80!black} $i \in [0 \, ..\, N-1]$ }{
        \tcp{\color{green!50!black} \protect\cite[Equations~(28) to~(31) and (50) to~(53)]{belcour2018}}
        $r_{0i}, r_{i0}, t_{0i}, t_{i0} = \mbox{UpdateEnergy()}$\;
        $\sigma^{R}_{0i}, \sigma^{R}_{i0}, \sigma^{T}_{0i}, \sigma^{T}_{i0} = \mbox{UpdateVariance()}$\;

        \BlankLine
        \tcp{\color{green!50!black} A new lobe is added to the list of BRDF lobes}
        \tcp{\color{green!50!black} with $ m = \nicefrac{r_{i0} r_{i}}{1 - r_{i0} r_{i}} $}
        ${\mbox{AddLobe}\left( {t_{0i} r_{i} t_{i0} \over 1 - r_{i0}r_{i}},\, \omega_r,\, {\scriptstyle \sigma^{T}_{i0} + j_{i0}\left[ \sigma^{T}_{0i} + \sigma^R_i \times m\times\left(\sigma^R_{i} + \sigma^R_{i0}\right) \right]} \right) }$\;

        \BlankLine
        \tcp{\color{green!50!black} Update scaling Jacobian}
        $j_{i0} = j_{i0} \times j_i$\; 
    }

    \BlankLine
    // {\color{green!50!black} \emph{Add the dieletric coating of the rough coated Lambertian} } \\
    ${\color{red!80!black} \mbox{AddLobe}\left( {t_{0i} r_{N} t_{i0} \over 1 - r_{i0}r_{N}},\, \omega_r,\, {\scriptstyle \sigma^{T}_{i0} + j_{i0}\left[ \sigma^{T}_{0i} + \sigma^R_N \times m\times\left(\sigma^R_{N} + \sigma^R_{i0}\right) \right]} \right) }$\;

    \BlankLine
    // {\color{green!50!black} \emph{Bottom-to-top loop to evaluate Lambertian lobe} } \\
    ${\color{red!80!black}r_{iN} = t_{0i} \times EvalReflectance()}$\tcp*{\color{green!50!black} \protect{Equation~(9)}}
    ${\color{red!80!black}\sigma_{iN} = EvalVariance()}$\tcp*{\color{green!50!black} \protect{Equation~(11)}}

    \BlankLine
    {\color{red!80!black}
        \For{$i \in [N-2 \, ..\, 0]$}{
            $r_{iN} = \nicefrac{ \left[ r_{iN} \, t_{i} \right] }{\left[ 1 - r_{i} r_{i+1} \right]}$\;
            $\sigma_{iN} = j_{i} \, \sigma_{iN} + \sigma^T_{i} + j_{i} \left( \sigma^R_{i} + \sigma^R_{i+1}  \right) \left( \dfrac{r_{i} r_{i+1}}{1 - r_{i} r_{i+1}} \right)$\;
            $\sigma_{iN} = \max(\sigma_{iN}, \sigma_{2+}(\eta_i, \alpha_i))$\;
        }
        $\mbox{AddLobe}(r_{iN}, \mathbf{n}, \sigma_{iN})$\;
    }
\end{algorithm}

\change{
\paragraph{A new operator.}  We decompose the operator for a rough coated Lambertian in two operators: one for the rough dielectric reflection and another for the multiple scattering componnent. This later part is mathematically defined as:
}
\begin{align}
    e       =\,& \rho_{2+} &\mbox{(as defined in Equation~\protect\ref{eqn:directional_energy})} \\
    \mu     =\,& 0 &\mbox{(aligned with the shading normal)} \\
    \sigma  =\,& \sigma_{2+} &\mbox{(tabulated, see Figure~\protect\ref{fig:tabulated}~(d))}
\end{align}
This creates an additional lobe (Figure~\ref{fig:update_layered_framework}, green lobe) that is centered around the shading normal. \change{Because of its different mean and variance, the transmission factor $t_{i0}$ is not valid for this lobe and we cannot apply the adding formula.} Hence, we add a specific case for the Lambertian lobe in the adding-doubling algorithm.

\paragraph{Modification of the adding-doubling algorithm.} To account for this additional lobe, we only modify a small part of the adding-doubling algorithm. The treatment of the dieletric interfaces still follows the implementation of Belcour~\shortcite{belcour2018} (see Algorithm~\ref{alg:new_adding_doubling} for a colored highlighted difference). Once the algorithm reaches the last dieletric interface, it adds the directly reflected lobe using the adding-doubling formula but handles the indirect lobe differently: since this lobe has a mean aligned with the shading normal, the transmission and reflection coefficients computed during the adding-doubling pass cannot be used as they were evaluated for the reflected direction. Those coefficients need to be reevaluated for all the interfaces using the mean and variance of the indirect lobe. We do so using a bottom-to-top loop; this is in fact equivalent to doing the adding-doubling for the transmittance when the layered structure is lit by the Lambertian layer (red changes of Algorithm~\ref{alg:new_adding_doubling}). \change{That is, we update reflection coefficient and variance from bottom to top using:
\begin{align}
    r_{iN} &= r_{i-1N} \dfrac{t_i}{1 - r_i r_{i+1}} \\
    \sigma_{Ni} &= j_i \sigma_{Ni-1} + \sigma_i^T + \sigma^T_{i} + j_{i} \left( \sigma^R_{i} + \sigma^R_{i+1}  \right) \left( \dfrac{r_{i} r_{i+1}}{1 - r_{i} r_{i+1}} \right)
\end{align}
using Equations~(29) and~(52) from the original paper.
}

\paragraph{High Variances.} The variance computed when transmitting through a dieletric interface should not surpass the variance of the diffuse transmission\footnote{$\sigma_{2+}(\eta,. \alpha)$ is the maximum achievable variance for unimodal distributions.} through a dielectric layer. We thus clamp the variance with it: $\sigma_t = \mbox{min}\left( \sigma_t, \sigma_{2+}(\eta, \alpha) \right).$

%
%
\section{implementation Details}

\paragraph{Dimensionality Reduction.} \change{In our implementation, we compress rough transmission $T_{01}$ from $4D$ to $2D + 1D + 1D$, diffuse transmission $\bar{T}_{10}$ and diffuse reflection $\bar{R}_{10}$ from 3D to $2D + 1D$  using 2 basis for absorption $\tau$ and 2 or 4 basis for roughness $\alpha$. For the latter, $2$ basis only requires $1$ RGBA texture while $4$ basis will require $2$. We show the quality of reconstruction in our supplemental material.}

\paragraph{Rough Coated Lambertian Secondary Lobe.} Our analysis shows that for $\eta > 1$, the secondary lobe for the rough coated Lambertian is visually close to a diffuse lobe. Using either an approximate GGX lobe (using the shading normal as the incident direction) or a diffuse lobe leads to similar appearance. However, this is not true for $\eta < 1$ where the variance of the lobe is bounded by the Total Internal Reflection (TIR) where a GGX lobe with the equivalent variance will match the discontinuity (see Figure~\ref{fig:profils}).

\section{Results}

\begin{figure}[t]
    \vspace{-7pt}
    \hspace{-15pt}
    \begin{tikzpicture}[font=\footnotesize]
        \node[inner sep=0] (ours_08_01) {
            \adjincludegraphics[height=2.7cm, frame, clip, trim=0 0 256 0]{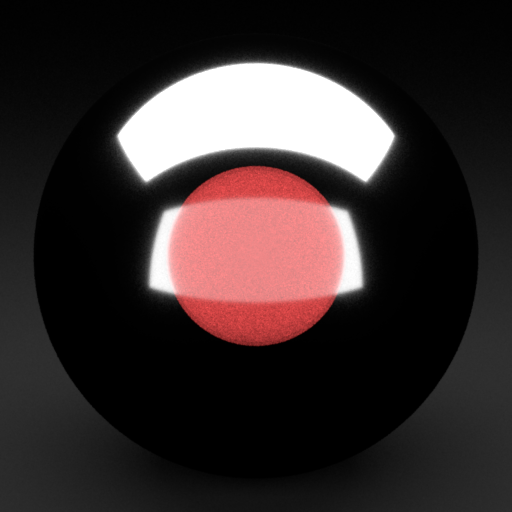}
        };
        \node[below=-2pt of ours_08_01, anchor=south] {\color{white} Ours };
        \node[inner sep=0, right=-1pt of ours_08_01, anchor=west] (ref_08_01) {
            \adjincludegraphics[height=2.7cm, frame, clip, trim=256 0 0 0]{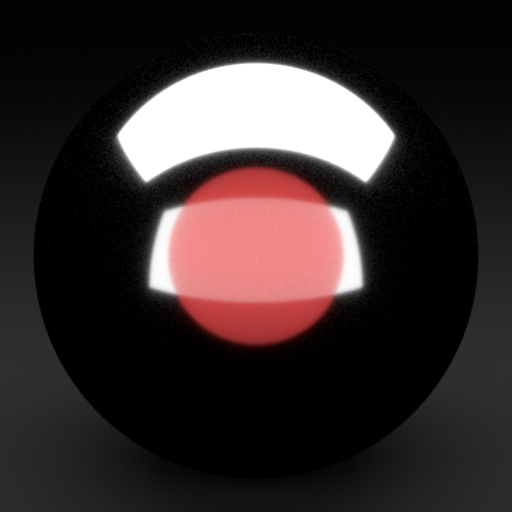}
        };
        \node[below=-2pt of ref_08_01, anchor=south] {\color{white} Reference };

        \node[inner sep=0, right=1pt of ref_08_01, anchor=west] (ours_15_01) {
            \adjincludegraphics[height=2.7cm, frame, clip, trim=0 0 256 0]{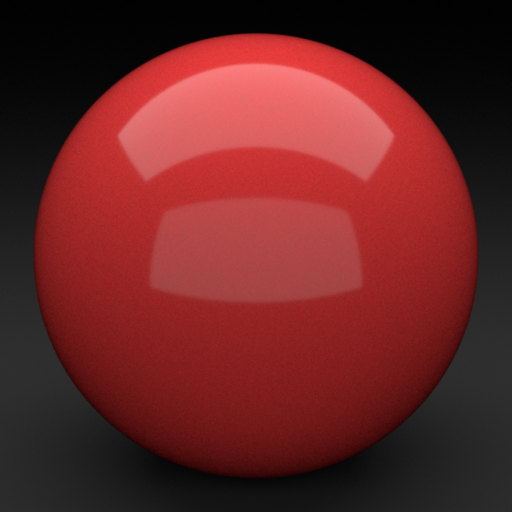}
        };
        \node[below=-2pt of ours_15_01, anchor=south] {\color{white} Ours };
        \node[inner sep=0, right=-1pt of ours_15_01, anchor=west] (ref_15_01) {
            \adjincludegraphics[height=2.7cm, frame, clip, trim=256 0 0 0]{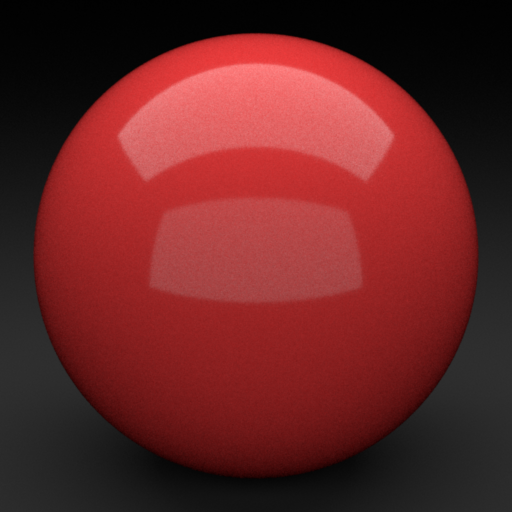}
        };
        \node[below=-2pt of ref_15_01, anchor=south] {\color{white} Reference };

        \node[inner sep=0, right=1pt of ref_15_01, anchor=west] (ours_30_01) {
            \adjincludegraphics[height=2.7cm, frame, clip, trim=0 0 256 0]{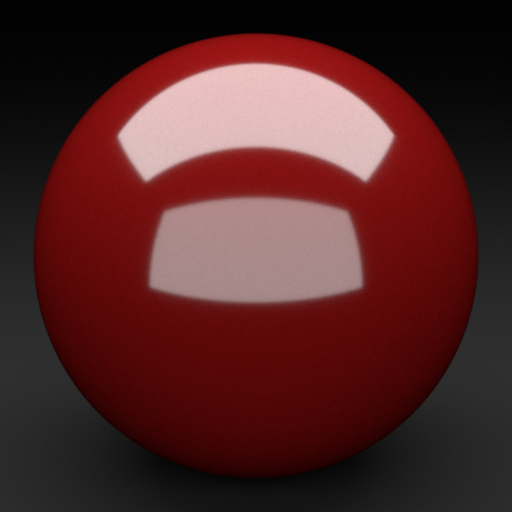}
        };
        \node[below=-2pt of ours_30_01, anchor=south] {\color{white} Ours };
        \node[inner sep=0, right=-1pt of ours_30_01, anchor=west] (ref_30_01) {
            \adjincludegraphics[height=2.7cm, frame, clip, trim=256 0 0 0]{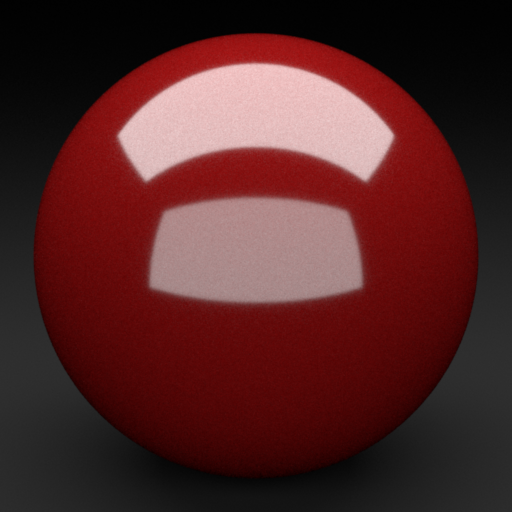}
        };
        \node[below=-2pt of ref_30_01, anchor=south] {\color{white} Reference };

        \node[inner sep=0, below=2pt of ours_08_01, anchor=north] (ours_08_3) {
            \adjincludegraphics[height=2.7cm, frame, clip, trim=0 0 256 0]{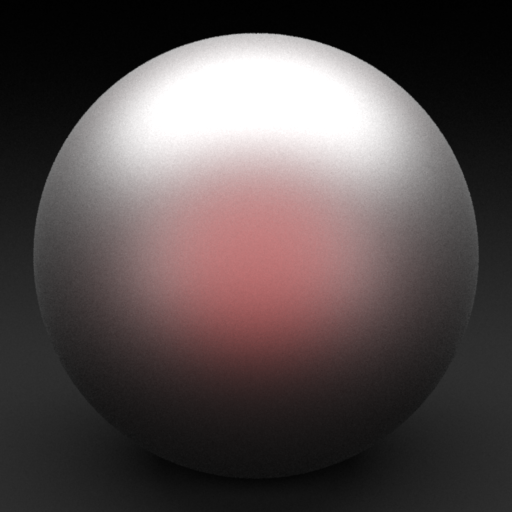}
        };
        \node[below=-2pt of ours_08_3, anchor=south] {\color{white} Ours };
        \node[inner sep=0, right=-1pt of ours_08_3, anchor=west] (ref_08_3) {
            \adjincludegraphics[height=2.7cm, frame, clip, trim=256 0 0 0]{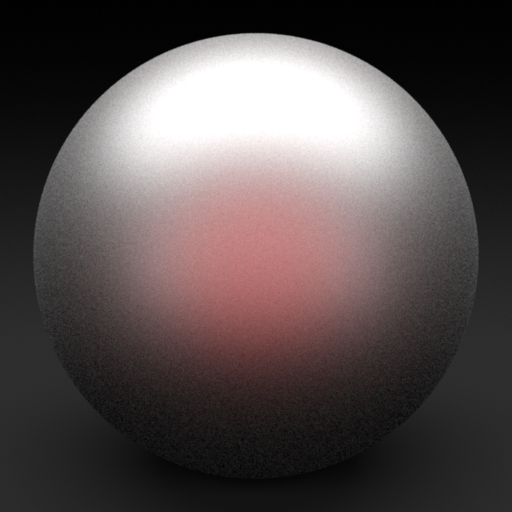}
        };
        \node[below=-2pt of ref_08_3, anchor=south] {\color{white} Reference };

        \node[inner sep=0, right=1pt of ref_08_3, anchor=west] (ours_15_3) {
            \adjincludegraphics[height=2.7cm, frame, clip, trim=0 0 256 0]{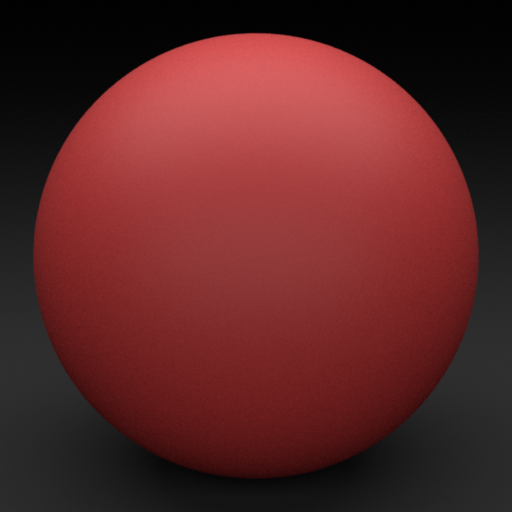}
        };
        \node[below=-2pt of ours_15_3, anchor=south] {\color{white} Ours };
        \node[inner sep=0, right=-1pt of ours_15_3, anchor=west] (ref_15_3) {
            \adjincludegraphics[height=2.7cm, frame, clip, trim=256 0 0 0]{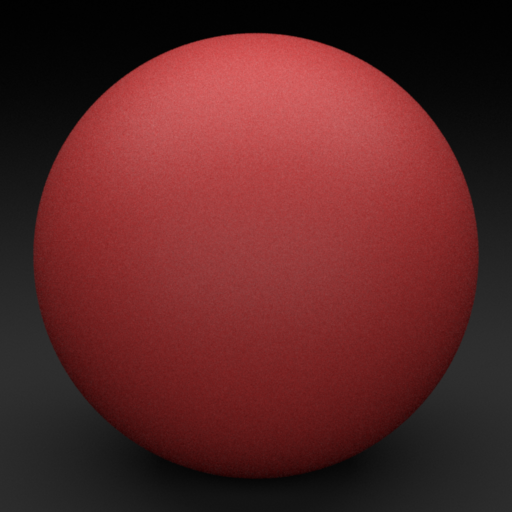}
        };
        \node[below=-2pt of ref_15_3, anchor=south] {\color{white} Reference };

        \node[inner sep=0, right=1pt of ref_15_3, anchor=west] (ours_30_3) {
            \adjincludegraphics[height=2.7cm, frame, clip, trim=0 0 256 0]{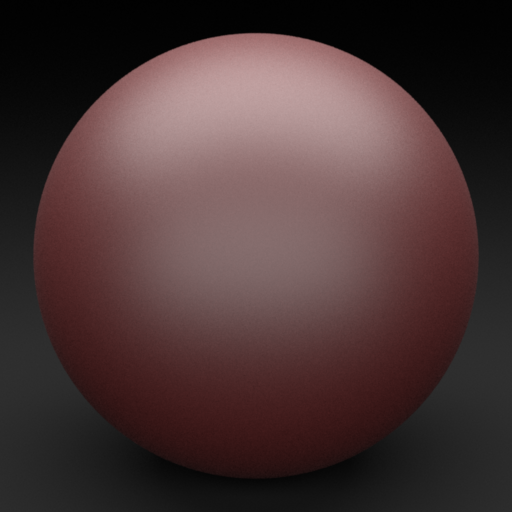}
        };
        \node[below=-2pt of ours_30_3, anchor=south] {\color{white} Ours };
        \node[inner sep=0, right=-1pt of ours_30_3, anchor=west] (ref_30_3) {
            \adjincludegraphics[height=2.7cm, frame, clip, trim=256 0 0 0]{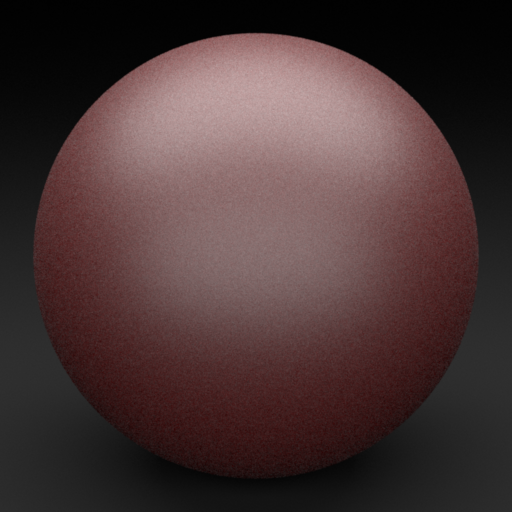}
        };
        \node[below=-2pt of ref_30_3, anchor=south] {\color{white} Reference };

        \node[left=1pt of ours_08_01, anchor=south, rotate=90] {$\alpha = 0.01$ };
        \node[left=1pt of ours_08_3,  anchor=south, rotate=90] {$\alpha = 0.3$ };

        \node at (0.6cm, 1.5cm) { $\eta = 0.4$ };
        \node at (3.5cm, 1.5cm) { $\eta = 1.5$ };
        \node at (6.2cm, 1.5cm) { $\eta = 3.0$ };
        
    \end{tikzpicture}
    \vspace{-7pt}
    \caption{
        Validation in Mitsuba.
        \textmd{
            We validated our rough coated Lambertian model in Mitsuba with a custom material plugin. As shown here we visually match a stochastic reference.
        }
        \label{fig:validation}
        \vspace{0pt}
    }
\end{figure}

\begin{figure}[th!]
    \hspace{-15pt}
    \begin{tikzpicture}[font=\footnotesize]
        \node[inner sep=0] (ours_08_01) {
            \adjincludegraphics[height=2.7cm, frame,]{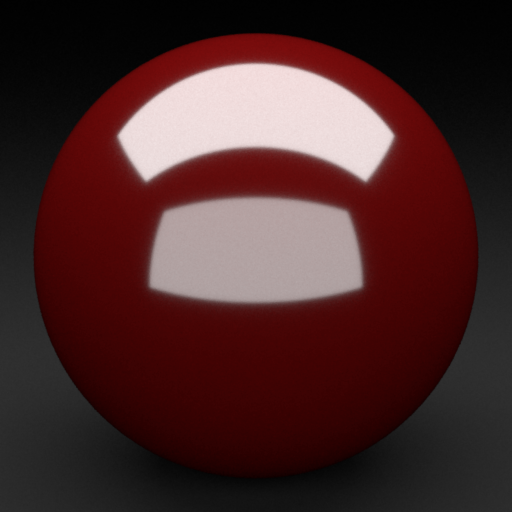}
        };

        \node[inner sep=0, right=1pt of ours_08_01, anchor=west] (ours_15_01) {
            \adjincludegraphics[height=2.7cm, frame]{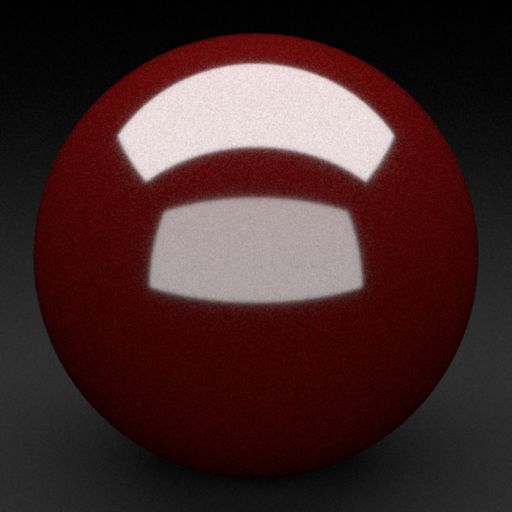}
        };

        \node[inner sep=0, right=1pt of ours_15_01, anchor=west] (ours_30_01) {
            \adjincludegraphics[height=2.7cm, frame]{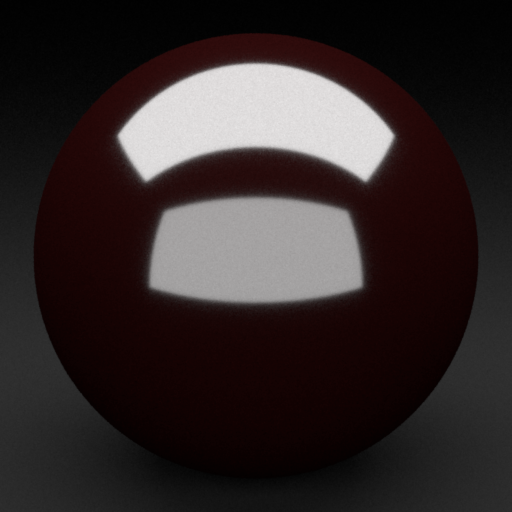}

        };

        \node[left=0pt of ours_08_01, anchor=south, rotate=90] { $\eta = 4.0$,  $\alpha = 0.01$};

        \node[inner sep=0, above=2pt of ours_08_01] (ours_08_02) {
            \adjincludegraphics[height=2.7cm, frame,]{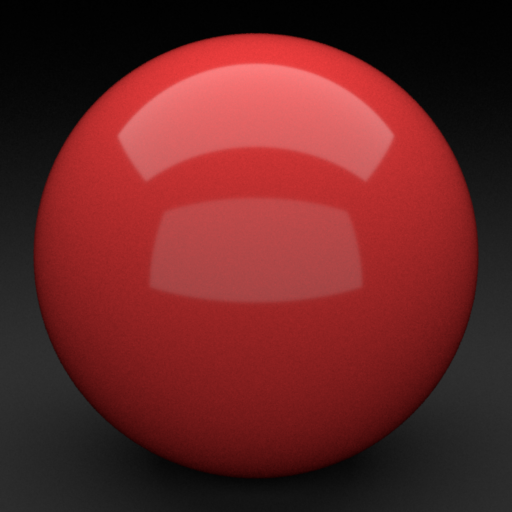}
        };
        \node[above=0pt of ours_08_02, anchor=south] { Ours };

        \node[inner sep=0, right=1pt of ours_08_02, anchor=west] (ours_15_02) {
            \adjincludegraphics[height=2.7cm, frame]{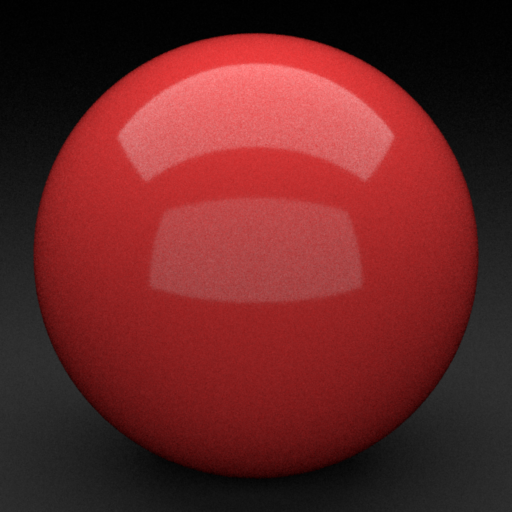}
        };
        \node[above=0pt of ours_15_02, anchor=south] { Reference };

        \node[inner sep=0, right=1pt of ours_15_02, anchor=west] (ours_30_02) {
            \adjincludegraphics[height=2.7cm, frame]{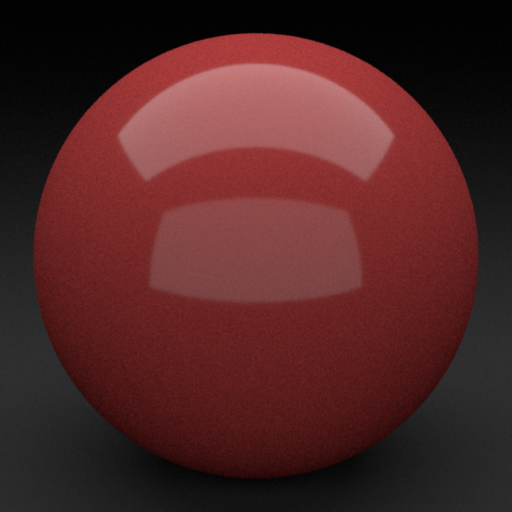}

        };
        \node[above=-1pt of ours_30_02, anchor=south] { W\&W~{\protect\shortcite{weidlich2007}} };

        \node[left=0pt of ours_08_02, anchor=south, rotate=90] { $\eta = 1.5$,  $\alpha = 0.01$};

    \end{tikzpicture}
    \vspace{-7pt}
    \caption{
        \change{
        Comparing to Weidlich and Wilkie~{\protect\shortcite{weidlich2007}}.
        \textmd{
            Our model (left) approximate multiple scattering within the coating and follows the reference (middle). On the contrary, the model of Weidlich and Wilkie~{\protect\shortcite{weidlich2007}} (right) lacks support for it.
        }}
        \label{fig:com_ww}
        \vspace{-20pt}
    }
\end{figure}

\paragraph{Validation in Mitsuba}
We validated the rough coated Lambertian model through a custom plugin in Mitsuba~\cite{mitsuba}. We compared this custom plugin to a stochastic reference capable of rendering arbitrary layered materials (similar to the one of Guo~\shortcite{guo2018}). Figure~\ref{fig:validation} showcases some of our unit tests. See our supplemental material for more results and source code. \change{Contrary to previous real-time compatible models, such as the one of Weidlich and Wilkie~\shortcite{weidlich2007}, our model accounts for multiple scattering (see Figure~\ref{fig:com_ww}). We validated that the configuration $\eta < 1$ generates plausible distributions (Figure~\ref{fig:validation}, left column). There, the inner red disc correspond to rays transmitted to the Lambertian base, while the outside of the disc correspond to pure reflection. While those a not realistic configurations, they are important when we layer different materials on top of each others. }

\paragraph{Real-time Prototype}
We implemented our rough coated Lambertian model as a GLSL fragment shader. This demo uses prefiltered Image Based Lighting~\cite{lagarde2014} for the environement maps. This shader visually matches the stochastic reference in Mitsuba as shown in Figure~\ref{fig:approximate_model}. In this prototype, we measured that shading all screen at 720p takes around $0.5$ms per frame on a RTX 2070. Please refer to our video for more details.

\begin{figure}[bh!]
    \begin{tikzpicture}[font=\footnotesize]
        \node[inner sep=0] (gl) {
            \includegraphics[width=0.5\linewidth, frame, clip, trim=0 100 640 100]{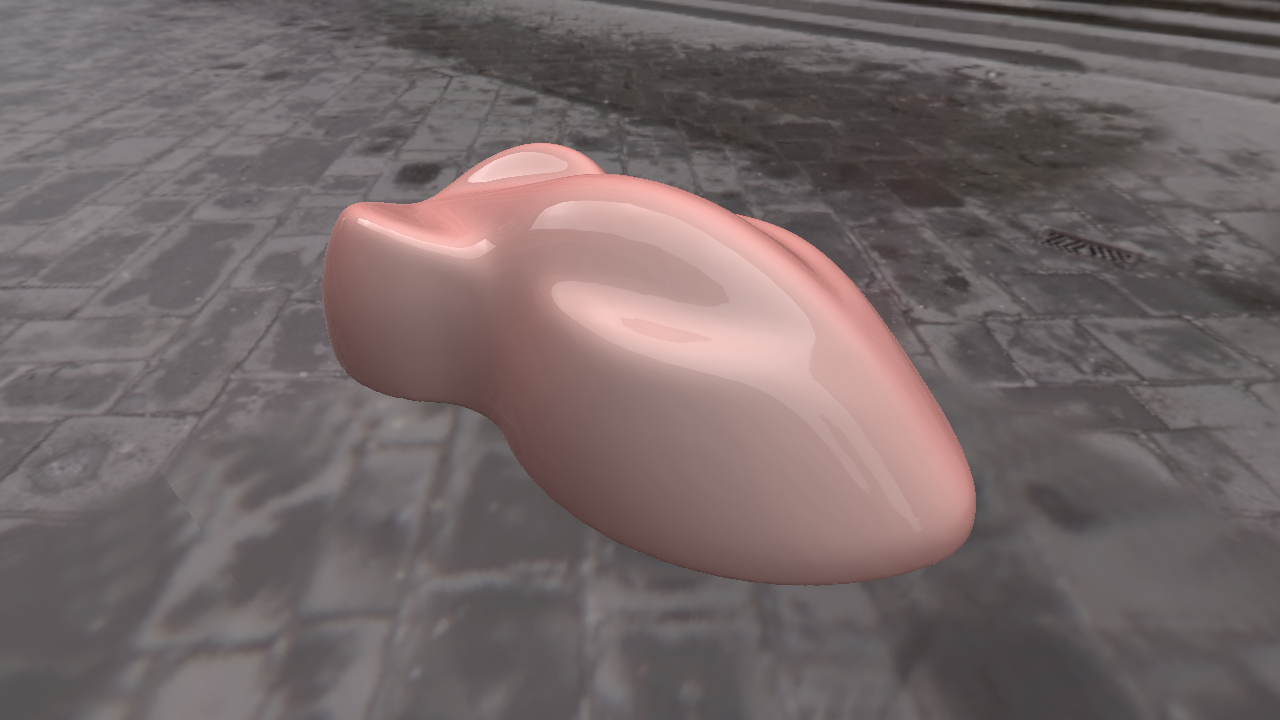}
        };
        \node[inner sep=0, right=-1pt of gl, anchor=west] (ref) {
            \includegraphics[width=0.5\linewidth, frame, clip, trim=640 100 0 100]{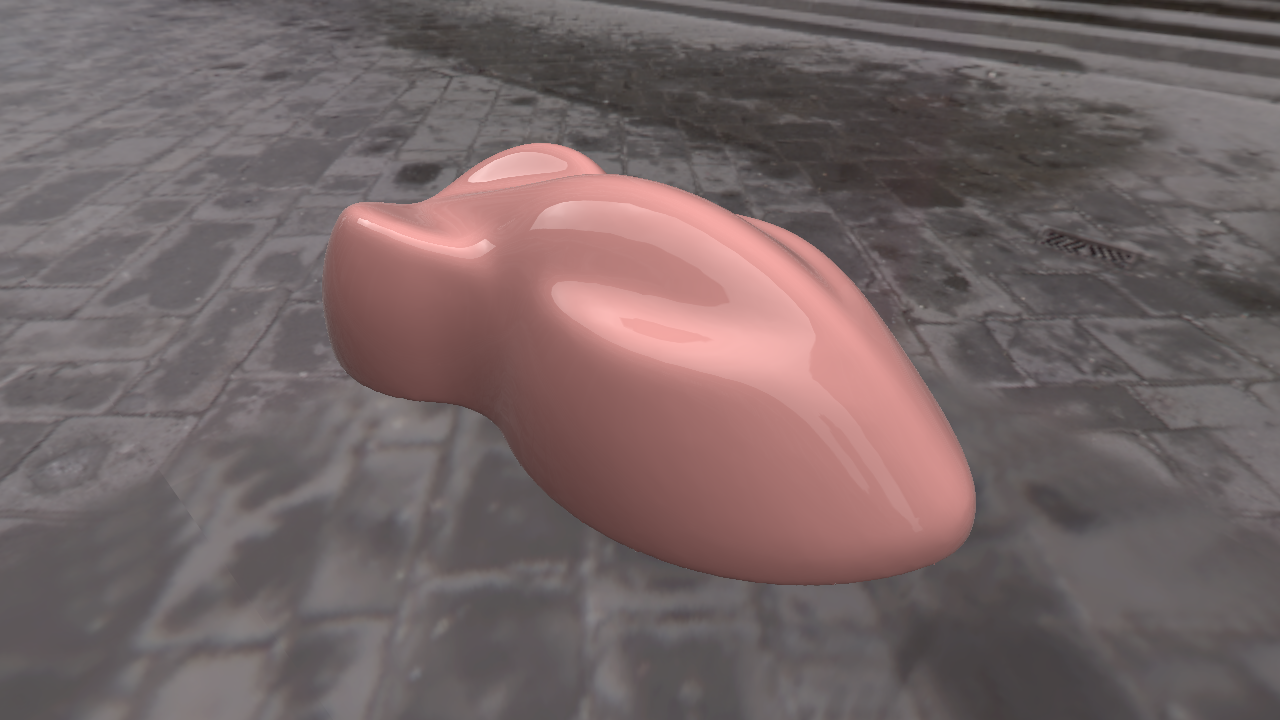}
        };
        \node[right=0pt of gl.south west, minimum width=0.5\linewidth, anchor=south west] { \textcolor{white}{Absorption+Albedo} };
        \node[right=0pt of ref.south east, minimum width=0.5\linewidth, anchor=south east] { \textcolor{white}{Albedo only} };

        \node[right=0pt of gl.north west, minimum width=0.5\linewidth, anchor=south west] (t1) { $\tau=\left[1.0, 0.64, 0.62\right]$};
        \node[above=-4pt of t1, minimum width=0.5\linewidth, anchor=south] { $\rho =\left[ 0.49, 1.0, 1.0\right]$ };

        \node[right=0pt of ref.north west, minimum width=0.5\linewidth, anchor=south west] (t2) { $\rho =\left[ 1.0, 0.45, 0.43\right]$ };
        \node[above=-4pt of t2, minimum width=0.5\linewidth, anchor=south] { $\tau=\left[0.8, 0.8, 0.8\right]$ };
    \end{tikzpicture}
    \vspace{-15pt}
    \caption{
        Unity prototype.
        \textmd{
            \change{We implemented our shading model in the Unity game engine. There, an artist can interactively change the appearance.
            For example,} using the absorption $\tau$, we can add subtle goniochromatic effects (left) that cannot be reproduced using the albedo only $\tau$ (right) by matching the grazing angle color.
        }
        \label{fig:unity}
    }
\end{figure}

\paragraph{Rendering in Unity}
We ported this shader in Unity's High Definition Render Pipeline~\cite{lagarde2021} as a custom Forward pass. There, our model runs at interactive framerate and allows artists to freely change its parameters with textures. In Figure~\ref{fig:teaser}, we reproduce the appearance of coated ceramics and lacquered surfaces. We do so by texturing the albedo and the roughness (left) or the index of refraction (right). In Figure~\ref{fig:unity}, we display how absorption $\tau$ can be used to add goniochromatic effects that are not reproducible with albedo only.

\paragraph{Validation of Layered Materials}
We implemented the updated adding-doubling in the Mitsuba rendering engine. There, we reproduce the appearance of surfaces consisting of 2 dieletric layers on top of a Lambertian base. We compare our model to a stochastic reference in Figure~\ref{fig:layered} and show that our model is visually close to the reference.

\input{./figures/multilayers/multilayers.tex}

%
%
\input{figures/validation/profils.tex}

\section{Limitations}
\paragraph{Critial Angle} For very small roughnesses and $\eta < 1$, the critical angle generates a hard discontinuity in the resulting BRDF. Since we are approximating such distribution with a GGX lobe, we cannot reproduce it. This effect is mitigated by the reflectance and transmittance $R_{01}$ and $T_{01}$ but the quality of the reconstruction there will depend on the resolution of those tables.

\paragraph{Anisotropy \& Participating Media} We restricted our model to rough isotropic dieletric coatings. However, adding an additional roughness dimension to the precomputed tables would not change the core of our method. We hypothetize that we could reduce this dimension using a few basis components as well. Another restriction is that we did not consider multiple scattering within the medium. Such transport requires to track many lobe directions~\cite{randrianandrasana2021} that negatively impact performances.

\section{Conclusion}
We presented a new shading model to render, in real-time, coated ceramics-like surfaces consisting of a Lambertian base coated by a rough dieletric interface, separated by an absorbing medium. We build this model from the numerical study of the first three moments of light transport in the layered structure. We showed that our model consisting of two BRDF lobes accurately reproduce the ground truth. We made this model compatible with real-time constraints by compressing the required tables to manageable sizes through dimensionality reduction. Leveraging those statistics, we added the support of Lambertian interfaces in the layered BRDF framework of Belcour~\shortcite{belcour2018}. This permits to increase the gamut of physically based appearances in real-time.

\begin{acks}
    The authors thanks Jonathan Dupuy for proof-reading the paper as well as Pascal Barla and M\'egane Bati for early discussions.
\end{acks}
\newpage

\bibliographystyle{ACM-Reference-Format}
\bibliography{bibliography}

\end{document}